\documentclass[twocolumn,showpacs,preprintnumbers,nofootinbib]{revtex4}

\usepackage{epsfig}
\usepackage{color}
\usepackage{ulem}
\usepackage{amsmath}
\usepackage{amssymb}
\usepackage{sidecap}

\newcommand{\dd}{\textrm d}

\begin{document}

\title{A critical analysis of the theoretical scheme to evaluate 
  photoelectron spectra}

\author{P.M. Dinh$^{1,2}$, P. Romaniello$^{1,3}$,
P.-G. Reinhard$^{4}$, and E. Suraud$^{1,2}$}
 
\affiliation{1) CNRS, LPT (IRSAMC) F-31062 Toulouse, France\\
2) Universit\'e de Toulouse, UPS,  Laboratoire de Physique
  Th\'{e}orique (IRSAMC) F-31062 Toulouse, France\\
3) European Theoretical Spectroscopy Facility (ETSF)\\
4) Institut f{\"{u}}r Theoretische Physik,
  Universit{\"{a}}t Erlangen, D-91058 Erlangen, Germany}

\begin{abstract}
We discuss in depth the validity and limitations of a
  theoretical scheme to evaluate 
photo-electron spectra (PES) through collecting the phase oscillations at a
given measuring point. Problems
appear if the laser pulse is still active when the first bunches of outgoing
flow reach the measuring point. This limits the simple scheme for evaluation
of PES to low and moderate laser intensities.  Using a model of free particle
plus dipole field, we develop a generalized scheme which is shown to
considerably improve the results for high intensities.
\end{abstract}

\pacs{33.60.+q, 33.80.Eh, 36.20.Kd}
                             
\maketitle

\section{Introduction}
Photo-electron spectra (PES) constitute a longstanding basic tool for
analyzing the electronic structure of atoms, molecules, or solids
\cite{Tur62a,Tur70aB,Gos83aB}. In the one-photon regime, PES provide
an image of the sequence of single-particle levels which are occupied
in the ground state.  This has been applied, e.g., already in the
early days of cluster physics to the electronic structure of cluster
anions to track the transition to bulk metal \cite{McH89}.
While this could be done with photon frequencies in the range of
visible light, the analysis of deeper levels requires higher
frequencies,  and hence one finds also UV \cite{Rab77aB} and X-ray PES \cite{Mou62aB}.  Nowadays, we
dispose of a great variety of coherent light sources in large ranges
of frequency, intensity and pulse length, such as the
very powerful and versatile free electron lasers
\cite{Gel10a,Bre10a} which now even allow for time-resolved studies of
deep lying core states of atoms embedded in a material
\cite{Pie08a}.  With the great availability of good light sources,
studies of PES are now found in all areas of molecular physics, from
atoms over simple molecules \cite{Tje90a} to complex systems as
clusters \cite{Hof01} and organic molecules \cite{Liu98a}.\\
There is hence a general need for a robust theoretical tool of analysis
of PES in various dynamical regimes. Traditional approaches to
compute PES rely on (multi-)photon perturbation theory \cite{Fai87}. A
few years ago, we have developed a technique to evaluate the PES
directly from numerical simulations of the electronic excitations on a
spatial grid representation (and also for a plane-wave representation)
\cite{Poh00,Poh01} allowing exploration of a rather wide range of
dynamical scenarios \cite{Poh00}.  The basic idea of the
  procedure is to use absorbing boundary conditions and to
  assume free particle propagation near the boundaries. The
technique has indeed been successfully applied to a variety
of clusters \cite{Rei03a,Fen08a}, for which significant electron
emission can be achieved with comparatively low laser
intensities. Atoms and molecules with larger ionization potentials
(IP), as e.g. carbon, require stronger fields instead and it
was found in first tests that the application of the recipe from
Ref.\ \cite{Poh00} can yield artefacts for the case of long pulses with high
intensity.  Therefore it is a timely task to re-inspect the method and to
develop improvements where found necessary. Such a critical analysis
of the evaluation of PES on a numerical grid is the aim of this paper.

We start in section \ref{sec:PES-raw} with a brief review of the
traditional recipe to evaluate PES. We show that, while this simple
recipe works well for low laser intensities, it produces unphysical
spectral features at high intensities, which call for a revision of
the method. We continue with a quick reminder of the gauge
freedom in describing the laser field. We discuss two choices of
gauge and explain the gauge transformation that connects the two
choices. The question which gauge is ideally to be used in the
evaluation of PES will be answered in the course of the further
sections. Afterwards, the presence of a strong laser field in
coexistence with the emitted electrons is dealt with by solving the
time-dependent Schr\"odinger equation for free electrons plus a
homogeneous laser field. This case still allows closed solutions which
is then used as a basic ingredient to define a refined recipe to
evaluate PES.  In section \ref{sec:Illustration}, the refined recipe
is then tested for the analytically solvable case of Gaussian wave
packets first and then numerically for a few realistic test cases.
%
%
\section{Theoretical evaluation of PES}
\label{sec:PES-raw}
In this section, we first give a brief review of the scheme for the
evaluation of PES introduced in \cite{Poh00,Poh01}.
After showing that it is not appropriate for high laser intensities,
we propose a generalized scheme that solves the problem. All
schemes discussed here and in the previous papers rely on a
mean-field description where each electron is associated with a
single-electron wave function $\psi_\alpha$. We will consider in the 
formal discussions one representative state and drop the index
$\alpha$ to simplify notations.

\subsection{PES scheme for free electron}
The PES are calculated in a very efficient manner exploiting the
features of the absorbing boundary conditions \cite{Poh00,Poh01}.
We give here a brief summary of the procedure and its motivation,
first for the simple case of 1D and then in 3D. It is to be noted that these
traditional recipes for evaluating the PES are deduced under the assumption of
free particle propagation near the bounds of the numerical box.
\subsubsection{The 1D case}

We choose a ``measuring point'' $z_\mathcal{M}$ far away from the center and
one or two grid points before the absorbing boundaries.  We record the
wave function $\psi(z,t)$ at the measuring point $z_\mathcal{M}$ during the time
evolution. This delivers the raw information from which we extract the PES.
In order to develop an appropriate recipe, we now need a few formal
considerations. 

We assume that the mean-field is negligible at  $z_\mathcal{M}$ such that
we encounter there free particle dynamics 
which is governed by 
the one-particle time-dependent Schr\"odinger equation
$\mathrm i\partial_t\psi(z,t)=-\frac{\nabla^2}{2}\psi(z,t)$. The
solution is
\begin{equation}
\psi(z,t)=\int \frac{\ \mathrm dk}{\sqrt{2\pi}}
\, g(k)\, e^{\mathrm{i}(kz-\omega t)} 
\label{Eqn:WF}
\end{equation}
with the condition that $k$ and $\omega$ satisfy the dispersion relation
$\omega=k^2/2$. Note that atomic units ($\hbar = m_e= e =4\pi\epsilon_0=
1$) are used here and throughout the paper.
The measurement of PES is practically a momentum analysis of the
outgoing wave packet at a remote side. In momentum space, the wave
function (\ref{Eqn:WF}) reads  
 \begin{equation}
   \widehat{\psi}(k,t)
   =
   \int\frac{dz}{\sqrt{2\pi}}e^{-\mathrm{i}kx}\psi(z,t)
   =
   \underbrace{\widehat{\psi}(k,0)}_{\propto g(k)}
   e^{-\mathrm{i}\omega t}
   \quad.
 \end{equation} 
 The probability to find an outgoing particle with momentum $k$ is thus
given by $|\widehat{\psi}(k,t)|^2=|\widehat{\psi}(k,0)|^2$, from which the PES can be obtained as
  
\begin{equation}
   \mathcal{Y}(E_\mathrm{kin})
   \propto
   \frac{1}{\sqrt{E_\mathrm{kin}}}\left|\widehat{\psi}(k,0)\right|^2
   \quad,
   \label{Eqn:Direct_yield}
 \end{equation}
where $E_\mathrm{kin}=k^2/2=\omega$ is the kinetic energy and we have taken into account the appropriate
energy density $\propto E^{-1/2}_{kin}$. This is what we call in this work the exact definition of PES developed close to the
 experimental procedure.

However, it is numerically extremely expensive to evaluate the PES by
Fourier transformation for a remote slot in coordinate space. To
overcome this difficulty, we obtain the expansion coefficient $g(k)$
from the Fourier transform in time-frequency space of the
wave function $\psi(z,t)$ collected at a measuring point
$z_\mathcal{M}$ near the absorbing boundary. We denote the Fourier
transformation in time by $\widetilde{\psi}$ to distinguish it from
the spatial Fourier transform $\widehat{\psi}$. This then reads~:
\begin{eqnarray}
\widetilde{\psi}(z_\mathcal{M},\omega') &=& \int \frac{\ \mathrm
  dt}{\sqrt{2\pi}} \, \psi(z_\mathcal{M},t) \, e^{\mathrm i\omega' t} \cr
&=&\int \frac{\ \mathrm dt}{\sqrt{2\pi}} \int \frac{\ \mathrm
  dk}{\sqrt{2\pi}} \, g(k) \, e^{\mathrm i(kz_\mathcal{M}-\omega t)}
e^{i\omega' t}\cr 
&=&\int \frac{\mathrm dk}{\sqrt{2\pi}} \, g(k) \, e^{\mathrm ikz_\mathcal{M}}
\delta(\omega-\omega') \ .
\label{Eqn:coefficient_1}
\end{eqnarray}
Since we collect $\psi(z_\mathcal{M},t)$ far away from the center and
close to the absorbing boundary, we can assume
that only outgoing waves will pass the measuring point
$z_\mathcal{M}$. We will hence have only positive wave vectors
\begin{equation}
  \omega
  =
  \frac{k^2}{2}
  \quad\longleftrightarrow\quad
  k=+\sqrt{2\omega}
  \quad,
\label{eq:define-k} 
\end{equation}
which allows us to approximate the last
line in (\ref{Eqn:coefficient_1}) as 
\begin{eqnarray}
\widetilde{\psi}(z_\mathcal{M},\omega') 
&\simeq&
\int_0^{\infty} \frac{\ \mathrm d\omega}{\sqrt{4\pi\omega}}
\, g(\omega) \,  
e^{i\, z_\mathcal{M} \sqrt{2\omega}} \, \delta(\omega-\omega') \cr
&=&\frac{g(\omega')}{\sqrt{4\pi\omega'}}e^{i\,
  z_\mathcal{M}\sqrt{2\omega'}}\ \theta(\omega') 
\ .
\label{Eqn:coefficient_2}
\end{eqnarray}
In the last line, $\theta$ stands for the Heaviside function.
Using (\ref{Eqn:coefficient_2}) and (\ref{Eqn:coefficient_1}), we
finally obtain
\begin{equation}
g(\omega)\simeq \sqrt{4\pi\omega} \, e^{-\mathrm i\,
  z_\mathcal{M}\sqrt{2\omega}} \ \widetilde{\psi}(z,\omega) \ ,
\label{Eqn:coefficient_3}
\end{equation}
where from now on, we will always consider $\omega>0$ and thus omit
the factor $\theta(\omega)$ in Eq.~(\ref{Eqn:coefficient_3}).
At measuring point $z_{\mathcal{M}}$, we can identify
$\widetilde{\psi}(z_{\mathcal{M}},\omega)$ with
$\widetilde{\psi}(z_{\mathcal{M}},E_{\rm kin})$ where $E_{\rm kin}$ is the kinetic
energy of the electron. This defines the PES yield
$\mathcal{Y}_{z_{\mathcal{M}}}(E_{\rm kin})$ as 
\begin{equation}
\label{eq:raw-recipe}
  \mathcal{Y}_{z_{\mathcal{M}}}(E_{\rm kin}) 
  \propto
  \sqrt{E_{\rm kin}}
  \left| \widetilde{\psi}(z_{\mathcal{M}},E_{\rm kin}) \right|^2 \ .
\end{equation}
We have checked the method in extensive 1D wave packet calculations
and compared it to a direct momentum decomposition of the outgoing
wave as given in Eq.~(\ref{Eqn:Direct_yield}). Both
methods yield the same results, while the above sketched frequency
analysis at a "measuring point" is orders of magnitude faster.  

\subsubsection{The 3D case}

As in 1D, we assume for the 3D case free
propagation and purely outgoing waves at the measuring point
$\mathbf{r}_\mathcal{M}$.
The solution  of the one-particle time-dependent Schr\"odinger
equation in free space reads
\begin{equation}
  \psi(\mathbf{r},t)
  =
  \int \frac{\ \mathrm d^3\mathbf{k}}{\sqrt{(2\pi)^3}}
  \, g(\mathbf{k}) \, e^{i(\mathbf{k}\cdot\mathbf{r}-\omega t)}
  \quad.
\end{equation}
Analogously as in Eq. (\ref{eq:define-k}) for the 1D case, we 
calculate the coefficients $g(\mathbf{k})$ from the time 
Fourier transform of $\psi(\mathbf{r},t)$, assuming that
only wave vectors $\mathbf{k}=k \,\mathbf e_\mathbf{r}$
with $k>0$ contribute, where  $\mathbf e_\mathbf{r}$ is 
the direction of the outgoing
radial wave. This yields
\begin{equation}
g(\omega,\Omega_{\mathbf{r}}) 
 \propto \,
\widetilde{\psi}(\mathbf{r},\omega) \ , 
\label{Eqn:coefficient_3D}
\end{equation}
where 
$\Omega_{\mathbf{r}}$ is the solid angle related to the $\mathbf{r}$
direction. Considering $\mathbf{r}_{\mathcal{M}}$ as the measuring point, we compute the PES in direction
$\Omega_{\mathbf{r}_\mathcal{M}}$ as
\begin{equation}
\mathcal{Y}_{\Omega_{\mathbf{r}_\mathcal{M}}}(E_{\rm kin}) \, \propto \, 
\left| \widetilde{\psi}(\mathbf{r}_{\mathcal{M}},E_{\rm kin})
\right|^2 \ .
\label{eq:raw3D}
\end{equation}

%
%
%
%
%
%
%


The above analysis yields the fully energy- and
angular-resolved PES. The angular averaged PES would then be attained by angular integration with proper solid angle weights. 
It has nevertheless to be emphasized that the resulting angular dependence in the
laboratory frame depends on the orientation of the molecule (or
cluster). Actual ensembles in the gas phase do not contain molecules in
well defined orientation but represent rather an equi-distribution of all
possible orientations.  A typical example are the many recent measurements of
angular-resolved PES in cluster physics see, e.g.,
\cite{Pin99,Wil04,Kos07a,Bar09,Kje10a}.  An appropriate orientation averaging
has to be performed before one can compare
$\mathcal{Y}_{\Omega_{\mathbf{r}_\mathcal{M}}}(E_{\rm kin})$ with
experimental data.  The practical procedures for that are outlined in
\cite{Wop10a,Wop10b}.  Orientation averaging, however, is
beyond the scope of the present paper and will be ignored in the
following. 

\subsubsection{Example of application and problem}

Although this simple scheme has been applied with success to a variety of
clusters, our recent attempts to compute the PES of small
covalent
molecules have raised some questions concerning its general applicability.
For high laser intensities, we find an unexpected shoulder in the PES at large
kinetic energies. These pattern were then also found for the
simple test case of Na$_n$ clusters with jellium background when going to
sufficiently large laser field strengths. This excludes particularities of
pseudo-potentials, local or non-local ones. The problem seems to reside in the
scheme to evaluate PES.

\begin{figure}[htbp]
\begin{center}
\epsfig{file=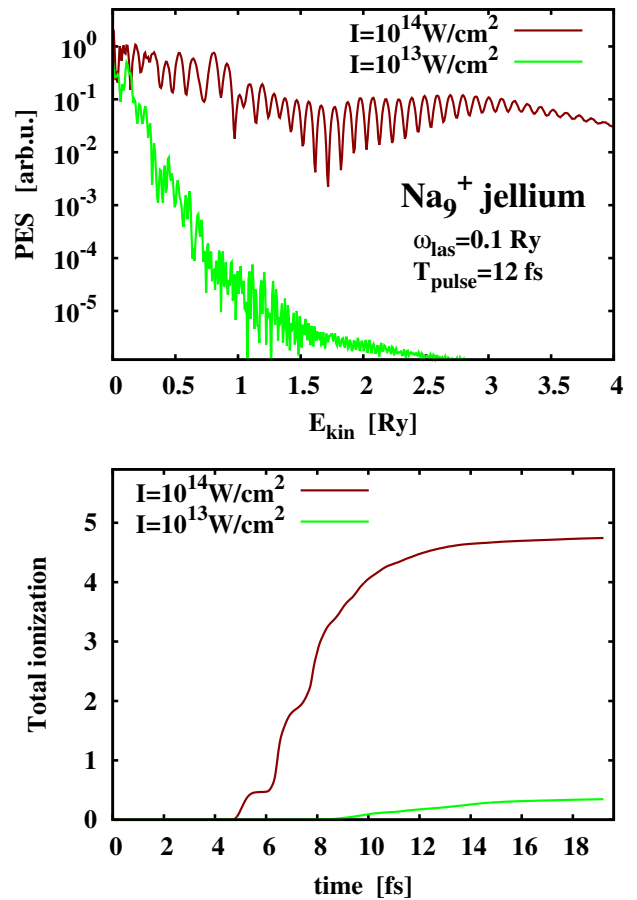,width=0.99\linewidth}
\caption{{
    Ionization properties of Na$_9^+$ with jellium background
    under the influence of laser pulses having
    frequency $\omega_\mathrm{las}=0.1$ Ry,
    pulse length $T_\mathrm{pulse}=12$ fs, and
    intensities $I=10^{13}$ W/cm$^2$ (green or light gray lines) and
    $I=10^{14}$ W/cm$^2$ (brown or dark gray lines), 
    computed in a cylindrical box of 176$\times$88 $a_0^2$ with
    spherical absorbing bounds covering at least 16 grid points.
    The laser field was given effectively in $v$-gauge.
    Bottom~: Time evolution of ionization.
    Left~: Photoelectron spectra
    $\mathcal{Y}_{\Omega_{\mathbf r_{\mathcal{M}}}}(E_{\rm kin})$, see Eq.\
    (\ref{eq:raw3D}). 
    }} 
\label{fig:Na9p-example}
\end{center}
\end{figure}

Figure \ref{fig:Na9p-example} 
demonstrates the problem for the case of
Na$_9^+$. The spherical jellium model is used for the ionic background.
Valence electrons are described by the time-dependent local-density
approximation (TDLDA) using the energy functional of 
\cite{Per92} and an average-density self-interaction  correction (ADSIC)
\cite{Leg02}.
Two cases are considered for comparison: one with still moderate
laser intensity, and another one in the high intensity regime.  In the
latter regime, more than half of the cluster's electrons are stripped
off, as shown in the bottom panel of Figure \ref{fig:Na9p-example}.
The top panel shows the PES for 
both cases.  For moderate intensity, we see the typical monotonous decrease.
The case of high intensity differs~: Up to $E_\mathrm{kin}\approx 1.5$ Ry,
we see the typical pattern of a monotonous decrease of the envelope with some
fluctuations. For $E_\mathrm{kin} >1.8$~Ry, however, we observe a new maximum, a
broad shoulder of high energy electrons; this is totally unexpected and most
probably unphysical.
In the following, we clarify the origin of
such a shoulder and we generalize the method to compute PES to
a wider range of laser intensities.

\subsection{Space gauge and velocity gauge}
\label{sec:gauge}
Before proceeding with a cure for the PES scheme illustrated above, we
review the choices of gauge for describing the laser field. The
external laser field is described in the limit of long wave lengths
and neglecting magnetic effects (electronic velocities are very small
at any time). The electron-laser interaction can be described either
by a time-dependent scalar potential $\Phi(\mathbf{r},t)$ or a time-dependent
vector potential $\mathbf{A}(\mathbf{r},t)$ in the Hamiltonian, which
in general reads  
\begin{equation}
H(\mathbf r,t) = \frac{1}{2} \left[\hat{\mathbf p} +
  \frac{{\mathbf A}(\mathbf{r},t)}{c} \right]^2 -
\Phi(\mathbf{r},t) \ .
\end{equation}
In the \emph{space gauge} ($x$-gauge), the laser field is described,
within the dipole approximation, by a scalar potential only:
\begin{subequations}
\label{eq:laser-x}
\begin{equation}
{\mathbf A}^{(x)}(\mathbf{r},t) = \mathbf 0, \qquad 
\Phi^{(x)}(\mathbf{r},t) = -E_0f(t) {z} \ ,
\label{Eqn:scalar}
\end{equation}
where the laser polarization is chosen along the $z$ axis, and $E_0$
is the maximum electric field of the laser.
$f(t)$ here denotes the temporal profile of the laser pulse, taken
as:
\begin{equation}
  f(t)
  =
   \sin(\omega_\mathrm{las}^{\mbox{}}t)
   \sin^2\!\left(\!\frac{\pi\, t}{T_\mathrm{pulse}}^{\mbox{}}\!\right)
   \theta(t)\ \theta(T_\mathrm{pulse}^{\mbox{}}\!-\!t) \ ,
\label{eq:laserprof}
\end{equation}
\end{subequations}
where $\omega_\mathrm{las}$ and $T_{\rm pulse}$ are respectively the
frequency and the duration of the laser pulse. $\theta$ denotes the
Heaviside function. 
We have chosen a smooth $\sin^2$ profile which
combines high spectral selectivity and finite extension.  The spectral
selectivity is needed because the pulse profile carries through to the
PES \cite{Poh01}. 

In the \emph{velocity gauge} ($v$-gauge) instead, the laser
field enters the Hamiltonian as a vector potential only:
\begin{subequations}
\label{eq:laser-v}
\begin{equation}
\label{Eqn:vector}
{\mathbf A}^{(v)}(\mathbf{r},t)=-cE_0 F(t)\, \mathbf{e}_z \ , \quad
\Phi^{(v)}(\mathbf{r},t) = 0 \ ,
\end{equation}
with
\begin{equation}
F(t)=  \int^t_{-\infty} \mathrm dt'\,f(t').
\end{equation}
\end{subequations}
Note that the vector potential in dipole approximation is spatially
constant.

Both gauges are equivalent. They are connected by a gauge
transformation which in general reads~:
\begin{subequations}
\begin{eqnarray}
  \Phi' &=& \Phi + \frac{\dot{\chi}}{c}
  \quad,
\\
  \mathbf{A}' &=& \mathbf{A} - \nabla{\chi}
  \quad,
\\
  \psi' &=&  \psi \, \exp{\left(\frac{i\chi}{c}\right)}
  \quad.
\end{eqnarray}
\end{subequations}
Let us assume that we take the $v$-gauge as a starting
point~: the laser field is described by the vector potential
$\mathbf{A}^{(v)}=-cE_0 F(t)\mathbf{e}_z$, whereas
$\Phi^{(v)}=0$.  If we want to gauge transform the vector
potential into the scalar potential, then
$\nabla\chi=\mathbf{A}^{(v)}$,
from where 
\begin{equation}
\chi=-cE_0 F(t)z \quad \Rightarrow \quad \Phi^{(x)} =
\frac{\dot{\chi}}{c}=-E_0 f(t)z \ .
\end{equation}
This finally gives 
\begin{equation}
\psi^{(x)} = \psi^{(v)} \exp{\left[-iE_0F(t)z\right]} \,
\label{eq:gaugepsi}
\end{equation}
which relates the wave function $\psi^{(x)}$ as computed by time
propagation under the action of the scalar potential
(\ref{Eqn:scalar}), and  $\psi^{(v)}$ for the case with the vector
potential (\ref{eq:laser-v}).
It is obvious that this phase factor is crucial in the present PES
analysis, while it can easily be ignored for local observables as,
e.g., dipole momentum or net ionization.

The phase transformation allows one to decouple the gauge used in the analysis
from that used in the time evolution. Assume that we solve the Schr\"odinger
equation using the potentials (\ref{eq:laser-v}). This immediately yields the
wave function $\psi^{(v)}$ in $v$-gauge. The same wave function could be
obtained by using (\ref{eq:laser-x}) and applying the reverse transformation
(\ref{eq:gaugepsi}) to recover $\psi^{(v)}$ from the $\psi^{(x)}$ as obtained
by propagation. In fact, this is the most efficient way to evaluate
$\psi^{(v)}$ as the operator (\ref{Eqn:scalar}) is purely local. It is to be
noted that the transformation (\ref{eq:gaugepsi}) is relevant for the
PES only if the outgoing wave reaches the measuring point $z_\mathcal{M}$
at a time where the signal $F(t)$ is still active.  For very large boxes, the
outgoing wave and the signal can avoid to coincide, in which case each gauge
yields the same result. However, typical box sizes used in practice
are not always that large.

The question is now which gauge is most suitable for evaluating PES.
The basic papers \cite{Poh00,Poh03a} state that the preferable choice
is the $v$-gauge. The argument there was a better convergence with box
size. We now speculate that this has to do with (non-)coincidence of
outgoing wave packet and laser signal. In the next section, this will
be scrutinized using a solvable model.


\subsection{PES scheme for free particles plus laser field}
\label{sec:freepluslaser}

In section \ref{sec:PES-raw}, we have deduced the recipe for evaluating PES
under the assumption that the potential is negligible at the measuring
point. Although this may hold for the typical mean field of a
system, we cannot easily exclude the presence of the laser field at this point
because the laser field is of extremely long range (wave length much larger
than system size in the dipole approximation that we use
here). Thus we extend the considerations to the case of a free 
particle plus laser field in dipole approximation which is still analytically
solvable. We confine the considerations to the 1D case. The extension to 3D is
straightforward. 

\subsubsection{Momentum-space wave function in $v$-gauge}
\label{sec:solve-v-gauge}

We first consider the Schr\"odinger equation of a free particle with the
external field (\ref{eq:laser-v}) in $v$-gauge
\begin{subequations}
\begin{eqnarray}
  \frac{1}{2} \big[ \hat{\mathbf p}-E_0F(t) \, \mathbf e_z \big]^2
  \psi^{(v)}(z,t) 
  &=&
 \mathrm i\, \partial_t\psi^{(v)}(z,t)
  \quad,
\label{eq:Hwavep}
\\
  \psi^{(v)}(z,0) &=& \psi^{(v)}_0(z)
  \quad.
\label{eq:psi0}
\end{eqnarray}
\end{subequations}
It is advantageous to expand the wave function at a given time
in momentum space
:
\begin{subequations}
\label{eq:FT-k}
\begin{eqnarray}
  \psi^{(v)}(k,t)
  &=&
  \int \frac{\ \dd z}{\sqrt{2\pi}} \, e^{- \mathrm ikz} \,
  \widehat{\psi}^{(v)}(z,t) \ ,
\label{eq:transform-k} \\
\widehat{\psi}^{(v)}(k,0) &=& \widehat{\psi}^{(v)}_0(k)\ .
\label{eq:FT-k0}
\end{eqnarray}
\end{subequations}
This yields the Schr\"odinger equation in the form
\begin{equation}
  \frac{1}{2}\big[ k-E_0F(t) \big]^2\, \widehat{\psi}^{(v)}(k,t)
  =
  \mathrm i\, \partial_t\widehat{\psi}^{(v)}(k,t) \ ,
\end{equation}
which solution reads~:
\begin{equation}
  \widehat{\psi}^{(v)}(k,t)
  =
  \exp{\left(
    -\frac{\mathrm i}{2}
    \int_{0}^t \mathrm dt' \, \big[ k-E_0F(t') \big]^2 
  \right)}\widehat{\psi}^{(v)}_0(k) \ .
\end{equation}
We rewrite the latter expression as
\begin{subequations}
\label{eq:FT-psi-v}
\begin{equation}
\label{eq:solve-k}
  \widehat{\psi}^{(v)}(k,t)
  =
  \exp{\big(
    -\mathrm i\omega t
    +\mathrm ik\delta q
    -\mathrm i\delta\Omega
   \big)}\, \widehat{\psi}^{(v)}_0(k)
   \end{equation}
   with
\begin{eqnarray}
  \omega
  &=&
  \frac{k^2}{2}
  \quad,
\label{eq:solve-k-w}
\\
  \delta q(t)
  &=&
  E_0 \int_{0}^t\mathrm dt'\,F(t')
  \quad,
\label{eq:delq}\\
  \delta\Omega(t)
  &=&
  \frac{E_0^2}{2}\int_{0}^t\mathrm dt'\,F(t')^2
  \quad.
\label{eq:delOmega}
\end{eqnarray}
\end{subequations}
Eqs.~(\ref{eq:FT-psi-v}) constitute the solution when propagating the
wave packet in $v$-gauge.

\subsubsection{Momentum-space wave function in $x$-gauge}

The Schr\"odinger equation of the same situation in $x$-gauge reads
\begin{subequations}
\begin{eqnarray}
\label{Eqn:SE_x}
  \left[\frac{\hat{\mathbf p}^2}{2} \, +\, 
  E_0zf(t) \right] \, \psi^{(x)}(z,t) 
  &=&
\mathrm i \, \partial_t\psi^{(x)}(z,t)
  \quad,
\\
  \psi^{(x)}(z,0) &=& \psi^{(x)}_0(z)
  \quad,
\\
  f
  &=&
  \partial_t F
  \quad.
\end{eqnarray}
\end{subequations}
Note that at $t=0$, the laser field is off and therefore, the
wave function in $x$-gauge is the same as that in $v$-gauge. 
In the following, we hence indicate the wave function 
in $k$ space at $t=0$ indifferently as $\widehat{\psi_0}(k)$ since the
upper script $(x)$ or $(v)$ is irrelevant.
Thanks to the gauge transformation (\ref{eq:gaugepsi}), the solution
of (\ref{Eqn:SE_x}) can be easily obtained from the 
solution (\ref{eq:solve-k}) in $v$-gauge, that is~:
\begin{equation}
  \widehat{\psi}^{(x)}(k,t)
  =
  \widehat{\psi}^{(v)}\left(k+E_0F(t) , t\right)  
  \quad.
\label{eq:gaugepsi-k}
\end{equation}
The wave function in $x$-gauge thus becomes
\begin{subequations}
\label{eq:FT-psi-x}
\begin{equation}
\label{eq:solve-x}
  \widehat{\psi}^{(x)}(k,t)
  =
  \exp{\left(
    -\mathrm i\omega t
    +\mathrm i K\delta q
    -\mathrm i\delta\Omega
   \right)} \ \widehat{\psi}_0(k),
   \end{equation}
   with
   \begin{eqnarray}
   K
   &=&
   k+E_0F(t)
   \quad,
\label{eq:defK}
\\
  \omega
  &=&
  \frac{K^2}{2}
  \quad,
\end{eqnarray}
\end{subequations}
where $\delta q$ and $\delta\Omega$ as given in Eqs. (\ref{eq:delq})
and (\ref{eq:delOmega}) respectively. This form is considerably more
involved than the solution (\ref{eq:solve-k}) in $v$-gauge,
since $K$ explicitly depends on time $t$. We hence take this as a first formal indication that the preferred gauge to
evaluate PES should be the $v$-gauge.

\subsubsection{Reconstruction of the PES}
\label{sec:reconst}
The scheme to evaluate PES, as sketched in section \ref{sec:PES-raw}, collects
a wave function at a certain coordinate-space point $z_\mathcal{M}$ and
performs Fourier transformation in frequency. This yields, by virtue of
$\omega=k^2/2$, the outgoing wave function in momentum space from
which we then deduce the PES. In extension of the simple form
(\ref{eq:raw-recipe}), we now want to take care of the possible
coincidence of the laser field at the measuring point with the
by-passing outgoing wave. This setup has been solved in section
\ref{sec:solve-v-gauge}. We are now going to 
apply the PES analysis to the solution (\ref{eq:solve-k}) and deduce
how to modify the collected signal at $z_\mathcal{M}$ in order to
properly describe the underlying unperturbed wave packet
$\widehat{\psi}_0(k)$. 

The relation between the wave function in frequency space and that
in momentum space is given by
\begin{eqnarray}
\widetilde{\psi}(z,\omega) &=&
\int\!\frac{\dd t}{\sqrt{2\pi}}\,e^{i\omega t}\psi(z,t) \cr
  &=&
\int \frac{\dd k}{\sqrt{2\pi}}\,e^{ikz}
\int\!\frac{\dd t}{\sqrt{2\pi}}\,e^{i\omega t}\widehat{\psi}(k,t) \ .
\label{eq:z-w/k-t}
\end{eqnarray}
%
In the case where the laser field at the measuring point is
negligible, one can see from Eqs.~(\ref{eq:solve-k}) and
(\ref{eq:solve-k-w}) that the
solution in $k$-space shrinks to a simple
$\widehat{\psi}^{(v)}(k,t)=\exp{\left(-i\omega' t\right)}\,
\widehat{\psi}_0(k)$ with $\omega'=\frac{k^2}{2}$. Then the
integration over $t$ in (\ref{eq:z-w/k-t}) 
immediately produces a $\delta(\omega-\omega')$, and that over $k$,
using the fact that $k=\sqrt{2\omega'}$ close to the absorbing
boundary, yields $\widetilde{\psi}^{(v)}(z,\omega) \propto
\widehat{\psi}_0(k)$.  Therefore, 
the PES in Eq.\ (\ref{Eqn:Direct_yield}) can be directly evaluated from
$\widetilde{\psi}(z,\omega)$ (see Eq.\ (\ref{eq:raw-recipe})).

Now we proceed to the case with non-vanishing laser field at the measuring
point, once again in $v$-gauge.  Trying to apply the time-frequency
transformation directly to the $k$-space solution (\ref{eq:solve-k}) runs into
trouble due to the non-trivial time dependences induced by the 
factors $\delta q(t)$ and $\delta\Omega(t)$.  An obvious solution is simply to
counter-weight the disturbing phase factors by a phase-correction factor
$\exp{\left(-\mathrm ik\delta{q}+\mathrm{i}\delta\Omega\right)}$.  This can be
done even in coordinate space because these factors do not depend on the
position. Therefore we calculate the PES from 
a phase-augmented Fourier transform
$\widetilde{\psi}^\mathrm{(PA)}$, which is
now proportional to
\begin{subequations}
\begin{eqnarray}
\label{eq:PES-PC}
  \widetilde{\psi}^\mathrm{(PA)}(z_\mathcal{M},\omega)
  &\propto&
  \int \dd t\,e^{ \mathrm i\omega t}
  e^{\mathrm{i}\varphi}
  \psi^{(v)}(z_\mathcal{M},t)
  \quad,
\\
  \varphi
  &=&
  -k\delta q + \delta\Omega
  \quad.
\end{eqnarray}
\end{subequations}
This modified Fourier transform is then used as in Eqs. (\ref{eq:raw-recipe}) yielding the recipe for the 1D case
\begin{equation}
  \mathcal{Y}_{z_{\mathcal{M}}}(E_{\rm kin}) 
  \propto
  \sqrt{E_{\rm kin}}
  \left| \widetilde{\psi}^{\mathrm{(PA)}}(z_{\mathcal{M}},E_{\rm kin})
  \right|^2 \ . 
\label{eq:PA-recipe}
\end{equation}
The generalization for the 3D case proceeds analogously.

The new move in this generalized evaluation is to augment the original recipe
by the phase-correction factor $e^{\mathrm{i}\varphi}$. This should allow the
application of the recipe in a wider range of laser intensities and time
profiles. The factor becomes negligible for weak laser field, or for fields
which do not interfere in time with the emitted particle flow. We shall call
in the following the former recipe (\ref{eq:raw-recipe}) the ``raw'' recipe
and the generalized form (\ref{eq:PA-recipe}) the ``phase-augmented'' (PA)
recipe.  A word of caution
is in order. The additional phase involves the mere momentum
$k$. The recipe (\ref{eq:PES-PC}) identifies that with the frequency $\omega$
according to Eq.~(\ref{eq:define-k}). This is just as valid as the
Fourier transformation is selective in frequency. This may be at stake in
cases of very violent dynamics. 

We run into unsurmountable trouble if we try to develop a generalized scheme
on the basis of the solution (\ref{eq:solve-x}) in $x$-gauge
because here even 
the instantaneous momentum $K$, as given in Eq. (\ref{eq:defK}), depends on
time. This is another strong indication that the $v$-gauge is the appropriate
starting point for evaluating the PES (as was already argued in
\cite{Poh00,Poh03a}). We will confirm this suspicion in the next section with a practical
example.

\section{Illustration}\label{sec:Illustration}
In this section, we show how the generalized scheme (\ref{eq:PES-PC})
performs. We first apply it to an exactly solvable model, and then to
more realistic cases.

\subsection{Analytical test case: Gaussian wave packets}
\label{sec:wave packet}

As a first test case, we consider a Gaussian wave packet whose
propagation can be 
analytically described, even in the presence of a (homogeneous) laser field.
The analytical solution, carried forth to far distance and beyond the lifetime
of the laser field, allows a direct evaluation of the PES by filtering the
momentum components by Fourier transformation from coordinate to momentum
space. This ``direct'' analysis exactly corresponds to the experimental
procedure, see Eq.\ (\ref{Eqn:Direct_yield}).
Thus we can
test the PES analysis in frequency space  against these ``exact''
results. 

\subsubsection{Gaussian wave packet in a laser field}

We consider a one dimensional system with a Hamiltonian consisting of the free kinetic energy plus the external laser field, as it was
discussed in section \ref{sec:freepluslaser}. The wave
function of the system at initial time is taken as the following
Gaussian function:
\begin{equation}
\psi(z,0)  =
\frac{1}{(\pi\mu_0)^{1/4}}\;
\exp{\left(\mathrm ip_0z-\frac{(z-q_0)^2}{2\mu_0} - \mathrm i \Omega_0 \right)}
\ .
\label{Eqn:initial_GW}
\end{equation}
The Fourier transform in $k$ space of (\ref{Eqn:initial_GW}) reads
\begin{equation}
  \widehat{\psi}_0(k)
  =
  \left(\frac{\mu_0}{\pi} \right)^{1/4} \, e^{\mathrm i q_0(p_0-k) -
    \mathrm i \Omega_0}
\ 
\exp{\left(-\frac{\mu_0}{2}(k-p_0)^2\right)}
  \ .
\end{equation}
The solution of the time-dependent Schr\"odinger equation is detailed in
Appendix~\ref{app:model-x}, using the $x$-gauge. 
It reads:
\begin{eqnarray}
  \psi^{(x)}(z,t)
  &=&
  \left(\frac{\mu_0}{\pi({\mu_0}^2 + t^2)} \right)^{1/4}
  e^{\mathrm i\left( p_0 - E_0F(t) \right)z}
\cr
  &&\exp{\left(-\frac{[z-q(t)]^2}{2 (\mu_0 + \mathrm it) }
       - \mathrm i\Omega(t)\right)}
  \ ,
\label{eq:psi-x-gauge}
\end{eqnarray}
with
\begin{eqnarray}
  q(t)
  &=&
  q_0
  +
  p_0t
  -
 E_0\int_{0}^t \dd t' \, F(t') \ ,
\label{eq:solve-q}
\\
  \Omega(t)
  &=&
   \Omega_0
  +
  \frac{p_0^2}{2}t
  -
  p_0E_0\int_{0}^t \dd t' \, F(t')
\nonumber\\
  && \qquad +
  \frac{{E_0}^2}{2}\int_{0}^t \dd t'F^2(t')
  +
 \frac{1}{2} \mathrm{atan}\left(\frac{t}{\mu_0}\right) \ .
\end{eqnarray}
Using now the gauge transformation  (\ref{eq:gaugepsi}), the wave
function in $v$-gauge becomes
\begin{eqnarray}
  \psi^{(v)}(z,t)
  &=&
  \left(\frac{\mu_0}{\pi({\mu_0}^2 + t^2)} \right)^{1/4}
  e^{\mathrm i p_0z}
\cr
  &&\exp{\left(-\frac{[z-q(t)]^2}{2 (\mu_0 + \mathrm it) }
       - \mathrm i\Omega(t)\right)}
  \ .
\label{eq:psi-v-gauge}
\end{eqnarray}



\subsubsection{Exact evaluation of PES by spatial
 Fourier analysis}

An exact evaluation of the PES has to analyze the momentum
content of a wave packet at a place where free propagation is reached.
The momentum distribution computed from the wave packet 
$\psi^{(x)}$ by spatial Fourier transformation
becomes
\begin{eqnarray}
  \widehat{\psi}^{(x)}(k,t)
  &=&
\left(\frac{\mu_0}{\pi}\right)^{1/4} e^{\mathrm i
    \arg(\mu)/2} \ 
  \exp{\big( -\frac{\mu}{2} [k-p(t)]^2 \big)}
\nonumber\\
  && \qquad
   \exp{\big(\mathrm{i}[p(t)-k]q(t) - \mathrm{i}\Omega(t) \big)}
\end{eqnarray}
where $\arg(\mu)$ denotes the argument of $\mu$.
The momentum distribution and corresponding probability becomes in $v$-gauge
\begin{eqnarray}
  \widehat{\psi}^{(v)}(k,t)
  &=&
  \left(\frac{\mu_0}{\pi}\right)^{1/4} e^{\mathrm i
    \arg(\mu)/2} \ 
  \exp{\left(-\frac{\mu}{2}[k-p_0]^2\right)}
\nonumber\\
  && \qquad
   \exp{\big(\mathrm i[p_0-k]q(t) - \mathrm i\Omega(t) \big)} 
  \quad,
\nonumber
\\
  \left|\widehat{\psi}^{(v)}(k,t) \right|^2
  &=&
  \left(\frac{\mu_0}{\pi}\right)^{1/2}
  \exp{\left(-{\mu_0}[k-p_0]^2\right)}
  \quad.
\end{eqnarray}
The corresponding kinetic energy distribution is obtained by the
identification $k=+\sqrt{2E_\mathrm{kin}}$. One also has to account for
the appropriate energy density $\propto E_\mathrm{kin}^{-1/2}$. This
yields
\begin{eqnarray}
  \mathcal Y^{(v)}(E_\mathrm{kin})
  &\propto&
  \sqrt{\frac{1}{E_\mathrm{kin}}}
  \left|\widehat{\psi}^{(x)}(\sqrt{2E_\mathrm{kin}})\right|^2
\nonumber\\
  &=&
  \sqrt{\frac{1}{E_\mathrm{kin}}}
  \left(\frac{\mu_0}{\pi}\right)^{1/2}
  e^{-{\mu_0}(\sqrt{2E_\mathrm{kin}}-p_0)^2}
\label{eq:PES-xFFT}
\end{eqnarray}
This is then the exact energy distribution evaluated from
spatial Fourier transform.
\subsubsection{Evaluation with the phase-augmented PES scheme}
The final PES analysis for the above model relies on the wave functions at the
measuring point, i.e.  $\psi^{(x)}(z_\mathcal{M},t)$ and
$\psi^{(v)}(z_\mathcal{M},t)$ as given in Eqs. (\ref{eq:psi-v-gauge}) and
(\ref{eq:psi-x-gauge}), respectively.  
We use the laser profile
(\ref{eq:laserprof}).  
As an actual example, we consider the following
parameters: $p_0=1\,\mathrm{a}_0^{-1}$, $q_0=0$, $\Omega_0=0$,
$\omega_\mathrm{las}=0.11\,\mathrm{Ry}=1.4\,\mathrm{eV}$,
$T_\mathrm{pulse}=8000\,\mathrm{Ry}^{-1}=384\,\mathrm{fs}$,
$\mu_0=100\,\mathrm{a}_0$ and various field strengths $E_0$.  The measuring
point was taken at $z_\mathcal{M}=1000\,\mathrm{a}_0$ or
$2000\,\mathrm{a}_0$. The first choice yields an overlap between laser
pulse and flow signal at $z_\mathcal{M}$ while the second choice decouples
them.  In the following discussion, we quantify the field strength in terms of
the laser intensity.  The correspondence is, e.g.,
$I=10^{14}$~W/cm$^2\leftrightarrow E_0=0.109$~Ry/a$_0$,
$I=10^{12}$~W/cm$^2\leftrightarrow E_0=0.0109$~Ry/a$_0$,
$I=10^{10}$~W/cm$^2\leftrightarrow E_0=0.00109$~Ry/a$_0$.

\begin{figure}[htbp]
\begin{center}
\epsfig{figure=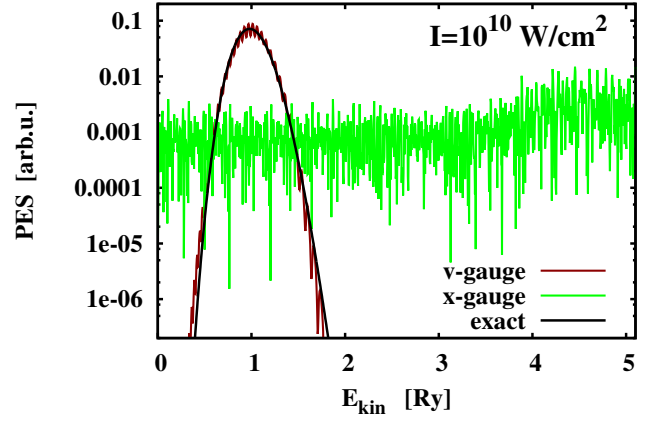,width=0.99\linewidth}
\caption{\label{fig:PES-wavepacket-gauge}
Photoelectron spectra from wave packet analysis
for a weak intensity $I=10^{10}$ W/cm$^2$.
We use the ``raw'' recipe (\ref{eq:raw-recipe}) and
compare analysis from the wave function in $v$- (brown or dark curve)
and $x$-gauge (green or light gray line).  
The exact result, that is the momentum
distribution of the wave packet as given in
Eq. (\ref{eq:PES-xFFT}) is also shown (black line).}
\end{center}
\end{figure}

Figure \ref{fig:PES-wavepacket-gauge} compares the evaluation of PES for
$x$-gauge versus $v$-gauge for a case of weak laser field where the phase
correction $\mathcal{\varphi}$ in Eq.~(\ref{eq:PES-PC}) is negligible
such that we 
can effectively use the ``raw recipe'' (\ref{eq:raw-recipe}). 
Laser pulse
length, wave packet, and measuring point have been chosen such that the wave
packet runs through the measuring point at a time where the laser is
still fully active.  It becomes obvious that the result from the wave function
in $x$-gauge is unphysical. The source of the problem lies in the
contribution
$\exp{\left(ip(t)z_\mathcal{M}\right)}$. 
As soon as the momentum $p=p(t)$ moves even slightly, the possibly large
$z_\mathcal{M}$ can amplify such a small oscillation and induce dramatic phase
oscillations which, in turn, produce a large contribution to the PES. This
contribution, however, must be unphysical because it sensitively depends on
the choice of the measuring point.  Clearly, the $v$-gauge is the preferred
choice for the evaluation of PES. This was already expected from the
analytical considerations in section \ref{sec:reconst}. We will henceforth
exclusively use the $v$-gauge.  Of course, a numerical solution of the
Schr\"odinger equation in coordinate space is often much simpler in
$x$-gauge. In such a case, one still can use the $x$-gauge for the solution and
then use the gauge transformation (\ref{eq:gaugepsi}) to bring this into
$v$-gauge. This is the path actually followed in section \ref{sec:naclust}.

\begin{figure}[htbp]
\begin{center}
\epsfig{figure=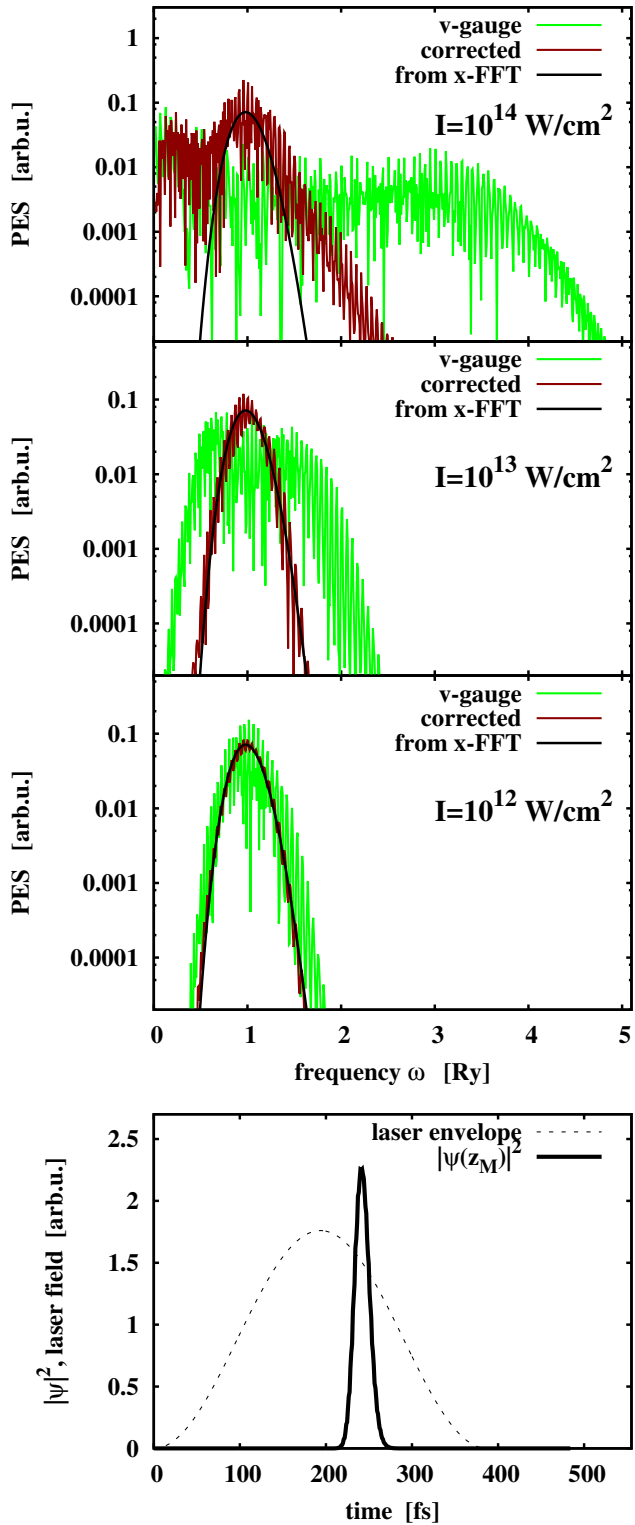,width=0.99\linewidth}
\caption{\label{fig:PES-wavepacket}
Photoelectron spectra from wave packet analysis.
Lower panel: The laser pulse envelope (dashed line) and the
probability density 
$|\psi(z_\mathcal{M},t)|^2$ at the measuring point (full line).
Upper panels: PES analysis at $z_\mathcal{M}$ in $v$-gauge from ``raw''
(green or light gray curve) and with phase augmented (PA) signal (brown
or dark gray line) for three
laser intensities as indicated. For comparison, the momentum
distribution of the wave packet as given in
Eq.~(\ref{eq:PES-xFFT}) is also shown (black curve).
}
\end{center}
\end{figure}

Results for higher intensities are collected in figure
\ref{fig:PES-wavepacket}.  
Here we use $v$-gauge throughout and compare the
``raw'' recipe (\ref{eq:raw-recipe}) with the PA recipe
(\ref{eq:PES-PC}).  The lowest panel shows the laser profile
(\ref{eq:laserprof}) together with the squared wave function
$|\psi(z_\mathcal{M},t)|^2$ at the measuring point. This signals a critical
situation where the laser is fully active while the wave packet is passing by
the measuring point. The three upper panels show results for different
intensities. The ``raw'' recipe still works acceptably well for the moderate
intensity $I=10^{12}$ W/cm$^2$, but becomes grossly misleading for higher
intensities. The generalized recipe (\ref{eq:PES-PC}) visibly improves
the performance. The results become reliable up to $I=10^{13}$
W/cm$^2$ and remain 
somehow qualitatively correct for the highest intensity. It is thus much
preferably to use the phase augmented form (\ref{eq:PES-PC}) for the
evaluation of the PES.

\begin{figure}[htbp]
\begin{center}
\epsfig{figure=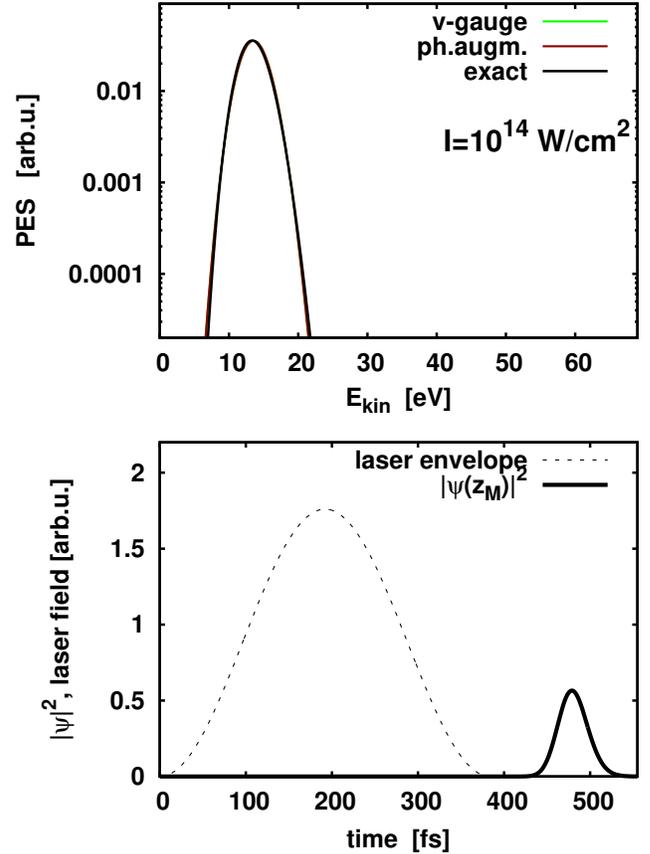,width=0.99\linewidth}
\caption{\label{fig:PES-wavepacket-no}
As in Figure \ref{fig:PES-wavepacket} but for a situation
where the laser pulse does not overlap with the wave packet at the
measuring point (see bottom panel).
Only the case of strongest laser intensity is shown.
}
\end{center}
\end{figure}

Figure \ref{fig:PES-wavepacket-no} shows results from a configuration where
the measuring point has been moved farther away to $z_\mathcal{M}=2000$ which
decouples the laser pulse from the wave packets signal.  
The case of highest
intensity $I=10^{14}$ W/cm$^2$ is shown. As expected, the
``raw'' recipe yields 
the same result as the PA one and both agree nicely with the
exact result. Moreover, the $x$-gauge (not shown here) yields
precisely the same result as the $v$-gauge. Thus all distinctions and
considerations are unnecessary in the case  that  laser pulse and
particle flow do not overlap.

\subsection{Realistic test cases}
\label{sec:naclust}

\begin{figure}[htbp]
\begin{center}
\epsfig{file=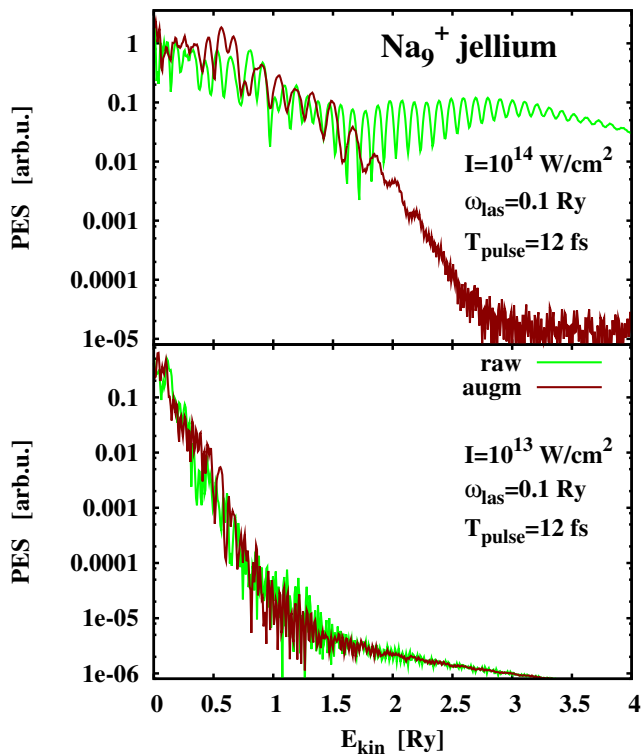,width=0.99\linewidth}
\caption{{
    Ionization properties for Na$_9^+$ with jellium background
    under the influence of a laser pulse having
    frequency $\omega_\mathrm{las}=0.1$ Ry, 
    pulse length $T_\mathrm{pulse}=12$ fs, and
    intensity as indicated, computed
    in a cylindrical box of 176$\times$88 $a_0^2$ with
    spherical absorbing bounds covering at least 16 grid points.
    The laser field was given effectively in $v$-gauge.
    The ``raw'' results (green or light gray curves) are given by
    Eqs. (\ref{eq:raw-recipe}), while  
    the PA results (brown or dark gray curves) are complemented by the
    phase factor (\ref{eq:PES-PC}).
    }} 
\label{fig:Na9p-example2}
\end{center}
\end{figure}

As a first realistic test case, we come back to the introductory
example of Figure~\ref{fig:Na9p-example}, namely the cluster Na$_9^+$ modeled
with spherical jellium background and
treated by time-dependent density functional theory using the energy
functional of \cite{Per92}.  We take the laser with a typical infrared
frequency and two rather large intensities.  
The axial symmetry of this and of the following example
allows us to perform the calculations in cylindrical coordinates.
Figure \ref{fig:Na9p-example2} 
shows the results for the ``raw'' and the PA recipes. 
The ``raw'' PES in the upper panel repeats the case of figure
\ref{fig:Na9p-example2} showing the obnoxious high energy shoulder. The PA PES
makes a dramatic difference. It shows a reasonable, almost monotonous,
decrease of the envelope. Only at about 2.5 Ry, the decrease turns
into a rather weak slope which may be unrealistic as we come here
into a region of very low yield where unwanted background may spoil the
analysis. The next lower intensity (lower panel) already shows reasonable
pattern with the ``raw'' recipe. The PA recipe brings some improvement as it
removes the glimpse of a shoulder at about 1.1 Ry. As already observed in the
analytic case of a Gaussian wave packet, lower intensities perform well
already with the ``raw'' recipe and the difference brought in from the PA
recipe is negligible.

\begin{figure}[htbp]
\begin{center}
\epsfig{file=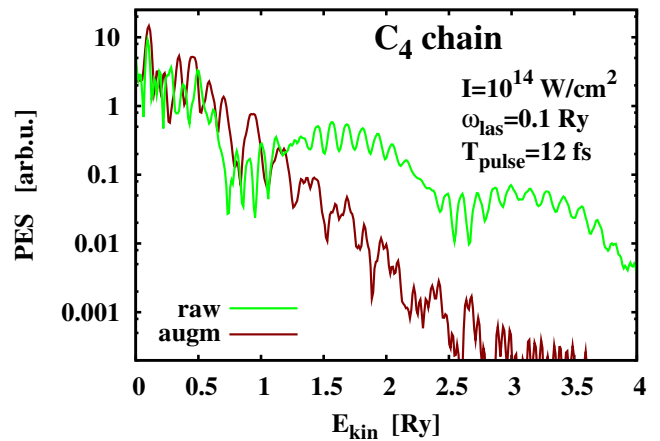,width=0.99\linewidth}
\caption{{
    Ionization properties for the C$_4$ chain
    under the influence of a laser pulse having
    frequency $\omega_\mathrm{las}=0.1$ Ry, 
    pulse length $T_\mathrm{pulse}=12$ fs, and
    intensity as indicated.
    The laser field was given effectively in $v$-gauge.
    The ``raw'' results (green of light gray curve) are given by
    Eqs. (\ref{eq:raw-recipe}), while the PA results (brown or dark gray
    line) are complemented by the phase factor (\ref{eq:PES-PC}).
    }} 
\label{fig:C4-example}
\end{center}
\end{figure}

Figure~\ref{fig:C4-example} shows a next test case results for the
C$_4$ chain, treated with non-local pseudo-potentials of Goedecker
type \cite{Goe96} and, again, the electronic energy-density functional
of \cite{Per92}.
The unphysical shoulders obtained with the ``raw'' recipe are even
more developed than in case of Na$_9^+$ and their successful removal
by the PA recipe is impressive. This is a
clear demonstration of the gain achieved by the PA recipe
(\ref{eq:PES-PC}). 
\section{Conclusion}
We have investigated the evaluation of photo-electron spectra (PES) by sampling the time evolution of the wave function at a fixed measuring
point.  The schemes were first analyzed using solvable models. The traditional
scheme was developed from the idea of a freely propagating outgoing wave. This
requires negligible potentials at the measuring point which, in turn, sets a
limit to an acceptable laser field strength. In order to extend the
applicability of the scheme for evaluating PES to stronger fields, we have
considered a model of a free particle plus an external laser field in dipole
approximation. This model is still analytically solvable and the analytical
solution allows us to deduce a more general scheme for the PES which consists
in augmenting the collected wave function by an appropriate phase factor
accounting for the time dependent laser field.  We have also investigated the
effect of gauge transformations on the results and pondered the question of
the most appropriate choice of gauge.  We have found that 
  the appropriate gauge for evaluating PES is clearly the $v$-gauge
  (velocity gauge).  Still, 
  the most practical way to produce the wave functions in $v$-gauge is
  to compute them first in $x$-gauge (space gauge) and to apply then
  the appropriate gauge transformation.
 
 The augmented scheme has been tested in
an analytical model of Gaussian wave packets and in a few realistic examples. A major result is that  
  the ``raw'' recipe for PES is valid for low and moderate laser fields
  (about $10^{12}$ W/cm$^2$ for the test case Na$_9^+$). It remains valid
  in case that the particle flow arrives at the measuring point after the
  laser pulse has died out (in this case for all intensities).
  The phase augmented evaluation of PES was shown to considerably extend
  the range of applicability. The gain was particularly dramatic for the
  example of the C$_4$ chain.
With the augmented
 recipe for computing PES  we have been able to check
  laser intensities up to $10^{14}$ W/cm$^2$
for our
  realistic test cases which all involved a low laser frequency of
  0.1 Ry. Higher laser frequencies reduce the effective intensity
  (Keldysh criterion). Thus even higher intensities may be used for
  higher frequencies.

After the successful tests shown here, the augmented recipe for evaluating
PES is ready for use in more demanding situations as they are molecular systems 
with large ionization potentials for which experimental data already exist as,
e.g., N$_2$, C$_{60}$ or typical organic molecules.

\appendix

\section{Solution of the Schr\"odinger equation for the wave packet model}
\label{app:model-x}
We provide the analytical solution of the wave packet propagation in
$x$-gauge. This is simpler and the $v$-gauge is regained easily by the
phase transformation (\ref{eq:gaugepsi}).  

The starting point is the Schr\"odinger equation as
given in Eq. (\ref{eq:Hwavep}).
The ansatz for the solution is given by
\begin{equation}
\label{eq:solve-analytic}
  \psi=
   \left(\frac{\mu_0}{\pi\,\mu^*(t)\mu(t)}\right)^{1/4}\;
  \exp{\!\left(ip(t)z\!-\!\frac{(z\!-\!q(t))^2}{2\mu(t)}
              -i\Omega(t)\!\right)}
 \end{equation}
with $\mu(t)=\mu_0+i\xi t$. The other time-dependent parameters are
determined substituting (\ref{eq:solve-analytic}) in the
time-dependent Schr\"odinger equation and comparing term by term. In
order to achieve this we first build the necessary derivatives:
\begin{eqnarray*}
  \frac{\mathrm i\partial_t\psi}{\psi}
  &=&
  -
  \frac{\mathrm i}{2}\frac{\xi^2t}{{\mu_0}^2 + \xi^2t^2}
  -
  \dot{p}z
  +
  \frac{i\dot{q}}{\mu}(z-q)
\\
  && \qquad
  -
  \frac{\xi}{2}\frac{(z-q)^2}{\mu^2}
  +
  \dot{\Omega}
\\
  \hat{p}\psi
  &=&
  \left(p+\frac{i}{\mu}(z-q)\right)\psi
\\
  \frac{\hat{p}^2\psi}{2\psi}
  &=&
  \frac{p^2}{2}
  +
  \frac{i p}{\mu}(z-q)
  +
  \frac{\mu_0 - \mathrm i \xi t}{2({\mu_0}^2 + \xi^2 t^2)}
  -
  \frac{1}{2}\frac{(z-q)^2}{\mu^2}
\\
  \frac{\Phi\psi}{\psi}
  &=&
  -E_0zf(t)
\end{eqnarray*}
Identifying term by term, we obtain the following equations
for the parameters
\begin{eqnarray*}
  (z-q)^2 
  &:&
 \xi = 1
\\
(z-q)
  &:&
  \dot{q}
  =
p
\\
  z
  &:&
  \dot{p}
  =
 -E_0\, f(t)
\\
  t
  &:&
  \xi^2 = \xi
\\
  z^0
  &:&
  \dot{\Omega}
  =
  \frac{p^2}{2}
  +
  \frac{\mu_0 - \mathrm i \xi t}{2 ({\mu_0}^2 + \xi^2 t^2)} + 
  \frac{\mathrm i \xi^2t}{2({\mu_0}^2 +\xi^2t^2)}
\end{eqnarray*}
from which one gets~:
\begin{subequations}
\label{eq:solve-analytic2}
\begin{eqnarray}
    \mu
  &=&
  \mu_0+it
  \quad,
\label{eq:sigma}
\\
  p
  &=&
  p_0  - E_0F(t)
  \quad,\quad
  F(t)=\int_{0}^t \mathrm dt' \, f(t')
  \quad,
\label{eq:solve-p}
\\  
  q
  &=&
  q_0
  +
  p_0t
  -
 E_0\int_{0}^t \mathrm dt' \, F(t')
  \quad,
\label{eq:solve-q2}
\\
  \Omega
  &=&
  \Omega_0
  +
  \frac{1}{2} \int_{0}^t \mathrm dt' 
\left( p^2 + \frac{\mu_0}{{\mu_0}^2 + t^2 }
  \right)
\nonumber\\
  &=&
  \Omega_0
  +
  \frac{{p_0}^2}{2}t
  -
  p_0E_0\int_{0}^t \mathrm dt' \, F(t')
\nonumber\\
  &+&
  \frac{{E_0}^2}{2}\int_{0}^t \mathrm dt' \, F^2(t')
  +
 \frac{1}{2} \mathrm{atan}\left( \frac{t}{\mu_0} \right) \quad.
\label{eq:solv-omega}
\end{eqnarray}
\end{subequations}

 \bibliography{cluster,add}

\end{document}